\documentclass[twocolumn]{aastex63}
\received{}
\revised{}
\accepted{}
\shorttitle{Magnetically driven winds from BNS merger remnants}
\shortauthors{Ciolfi R.}
\begin{document}

\title{Magnetically driven  baryon winds from binary neutron star merger remnants\\ and the blue kilonova of August 2017}

\correspondingauthor{Riccardo Ciolfi}
\email{riccardo.ciolfi@inaf.it}

\author[0000-0003-3140-8933]{Riccardo Ciolfi}
\affiliation{INAF, Osservatorio Astronomico di Padova, Vicolo dell'Osservatorio 5, I-35122 Padova, Italy}
\affiliation{INFN, Sezione di Padova, Via Francesco Marzolo 8, I-35131 Padova, Italy}

\author[0000-0002-2945-1142]{Jay Vijay Kalinani}
\affiliation{Universit\`a di Padova, Dipartimento di Fisica e Astronomia, Via Francesco Marzolo 8, I-35131 Padova, Italy}
\affiliation{INFN, Sezione di Padova, Via Francesco Marzolo 8, I-35131 Padova, Italy}



\begin{abstract}

\noindent The observation of a radioactively-powered kilonova associated with the first binary neutron star (BNS) merger detected in gravitational waves proved that these events are ideal sites for the production of heavy r-process elements. However, the physical origin of the ejected material responsible for the early (``blue'') and late (``red'') components of this kilonova is still debated. Here, we investigate the possibility that the early/blue kilonova originated from the magnetically driven baryon wind launched after merger by the metastable neutron star remnant. Exploiting a magnetized BNS merger simulation with over 250\,ms of post-merger evolution, we can follow for the first time the full mass ejection process up to its final decline. 
We find that the baryon wind carries $\simeq\!0.010\!-\!0.028\,M_\odot$ of unbound material, proving that the high mass estimated for the blue kilonova can be achieved. 
We also find expansion velocities of up to $\sim\!0.2\,c$, consistent with the lower end of the observational estimates, and we discuss possible effects neglected here that could further increase the final ejecta velocity.
Overall, our results show that the magnetically driven baryon wind represents a viable channel to explain the blue kilonova.

\end{abstract}



%%%%%%%%%%%%%%%%%%%%%%%%%%%%%%%
\section{Introduction} 
\label{sec:intro}

\noindent The first combined gravitational wave (GW) and electromagnetic (EM) observation of a binary neutron star (BNS) merger in August 2017 marked a major milestone in the investigation of these systems \citep{LVC-BNS,LVC-MMA, LVC-GRB}. Among a number of key discoveries, this event provided compelling evidence that BNS mergers can produce a copious amount of heavy r-process elements up to atomic mass numbers well above $A\!=\!140$ (e.g., \citealt{Kasen2017,Pian2017}), in agreement with decades of theoretical predictions \citep{Lattimer1974,Symbalisty1982}. 
Such confirmation was enabled by the detection and characterization of the UV/optical/IR signal AT\,2017gfo, which resulted fully consistent with being a kilonova, i.e.~a thermal transient powered by the radioactive decay of the heavy elements synthesized within the material ejected during and after merger \citep{Li1998,Metzger2010}.

Besides the overall consistency with a radioactively powered kilonova, however, the interpretation of AT\,2017gfo in terms of specific source properties remains challenging, not only due to uncertainties in the involved microphysics, but also because the observed signal may result from a complex combination of contributions from different ejecta components (e.g.,~\citealt{Metzger2019LRR,Siegel2019}). 
In particular, the analysis of the UV/optical/IR data showed that the simplest kilonova model based on isotropic ejecta with constant opacity and characteristic radial velocity does not provide a satisfactory description, suggesting instead the presence of at least two distinct ejecta components, respectively explaining an early ``blue kilonova'', peaking $\sim\!1$\,day after merger at UV/blue wavelengths, and a following ``red kilonova'', peaking $\sim\!1$\,week after merger in the IR band (e.g., \citealt{Kasen2017,Cowperthwaite2017,Pian2017,Villar2017}).
The physical origin and properties of these two components are still a matter of debate (e.g., \citealt{Metzger2018,Siegel2018,Kawaguchi2018,Nedora2019}), as well as the possibility of more than two distinct contributions (e.g., \citealt{Perego2017b,Villar2017}).

Here, we focus the attention on the origin of the blue kilonova, for which the inferred ejecta properties combine a relatively low opacity of $\simeq\!0.5$\,cm$^2$/g with a rather high mass of $\simeq\!0.015\!-\!0.025\,M_\odot$ and a velocity as high as $\simeq\!0.2\!-\!0.3\,c$ (e.g., \citealt{Kasen2017,Cowperthwaite2017,Pian2017,Villar2017}). 
Shock-driven dynamical ejecta can easily reach these velocities and part of the material could even maintain a low enough opacity, but the mass of the low opacity portion may be insufficient to fully explain the inferred value (e.g., \citealt{Bauswein2013,Hotokezaka2013,Radice2016}, but see also \citealt{Nedora2019}).
On the other hand, post-merger baryon winds launched by the accretion disk surrounding a newly formed black hole (BH) can accomodate very high masses and a wide range of opacities, but their typical velocity is limited to $\sim\!0.1\,c$ (e.g., \citealt{Siegel2018}).
A third option, somewhat intermediate, is the baryon wind launched by the (meta)stable massive neutron star (MNS) remnant prior to its eventual collapse to a BH (if any).
As pointed out, e.g., in \citet{Metzger2018}, the ongoing neutrino emission from the MNS can significantly raise the electron fraction of the wind material, suppressing the nucleosynthesis of the heaviest elements ($A\!>140$) and thus keeping the opacity low (e.g., \citealt{Perego2014}), and, at the same time, the wind velocity and mass flow rate can be significantly enhanced by the presence of a strong magnetic field.
\begin{figure*}[!t]
  \centering
  \includegraphics[width=0.999\linewidth]{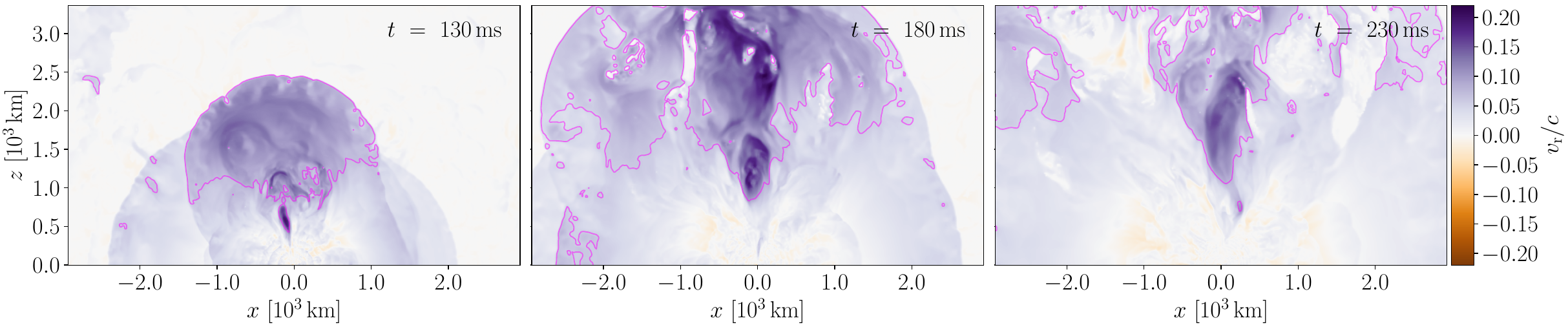}
  \caption{Meridional view of radial wind velocities at 130, 180, and 230\,ms after merger. The magenta contour lines indicate unbound matter (according to the geodesic criterion $-u_t\!>\!1$, see text).}
  \label{fig1}
\end{figure*}
While the magnetically enhanced MNS wind appears as a promising alternative, however, the possibility to reach high enough ejecta mass and velocity remains to be demonstrated in actual BNS merger simulations.  

In this Letter, we address the question on the viability of the magnetically driven baryon wind from the MNS remnant as an explanation for the blue kilonova in AT\,2017gfo. 
We do so by analysing the outcome of a general relativistic magnetohydrodynamics simulation of a BNS merger with over 250\,ms of MNS remnant evolution.\footnote{Some first results from this simulation were presented in \citet{Ciolfi2020a}, where the discussion was entirely focussed on the prospects of jet formation and the possible connection with short gamma-ray bursts.} 
This is to date the only simulation showing the emergence of a helical magnetic field structure and a collimated outflow along the MNS rotation axis as a natural consequence of the merger and post-merger dynamics\footnote{This is different from, e.g., \citet{Moesta2020}, where a strong, extended dipolar field aligned with the MNS spin is imposed by hand at an arbitrary time early after merger.} (where a similar outcome was previously obtained for accreting BH remnants; \citealt{Ruiz2016}). 
Our study does not include neutrino radiation nor a proper assessment of the ejecta composition and as such it is not aimed at producing kilonova lightcurves to be directly compared with the observations. 
Our aim is to show, through an example model, that a magnetically enhanced MNS wind can actually lead to an ejecta component as massive and fast as required by the observed blue kilonova. 

The very long post-merger evolution allows us, for the first time, to follow the specific mass ejection process associated with the magnetically driven wind until the mass flow rate has substantially declined, after which a further extrapolation in time becomes reliable. 
In turn, we can reconstruct the full time profile of the mass flow rate at a given distance, as well as the maximum ejecta mass attainable for a sufficiently long-lived MNS.
For the case at hand, we obtain a mass of the unbound wind material in the range $\simeq\!0.010\!-\!0.028\,M_\odot$, where most of the mass ejection (90\%) takes place between about 50 and 190\,ms after merger. This demonstrates that matching the mass requirement for the blue kilonova in AT\,2017gfo is possible.
Moreover, we find ejecta radial velocities distributed in the range $\simeq\!0.1\!-\!0.2\,c$ (with a maximum of $0.22\,c$), showing how the additional contribution of magnetic pressure can significantly enhance the velocity of the MNS wind (otherwise limited to $\lesssim0.1\,c$). 
Compared to the blue kilonova estimates, these velocities are consistent with the lower end of the inferred range. 
We also discuss various effects not included here that may further increase the ejecta velocities. 

Overall, our findings support the MNS wind as a viable source for the blue kilonova, thus encouraging further investigation in this direction.

%%%%%%%%%%%%%%%%%%%%%%%%%%%%%%%
\section{BNS model and numerical setup} 
\label{sec:setup}

\noindent The BNS merger under consideration has the same chirp mass as inferred for the GW170817 event \citep{LVC-170817properties} and a mass ratio of $q\!\simeq\!0.9$ (well inside the estimated range). NS matter is described via the piecewise-polytropic approximation of the APR4 equation of state (EOS) \citep{Akmal1998} (see also \citealt{Endrizzi2016}), which leads to a MNS remnant that survives without collapsing to a BH for the full extent of the simulation. 

Initial dipolar magnetic fields are imposed within the two NSs with maximum field strength of $5\times10^{15}$\,G and the total inital magnetic energy is $E_\mathrm{mag}\!\simeq\!4\times10^{47}$\,erg.
Such a high initial field strength allows us to reproduce the expected post-merger magnetization level ($E_\mathrm{mag}\!\sim\!10^{51}$\,erg) despite the fact that the early magnetic field amplification, in particular via the Kelvin-Helmholtz instability (e.g., \citealt{Kiuchi2015}), is not fully resolved. 
As discussed in more detail in \citet{Ciolfi2019} and \citet{Ciolfi2020a}, the magnetohydrodynamic evolution is rather well resolved starting from $\sim\!30-40$\,ms after merger, giving us confidence that our qualitative description of the relevant physical effects taking place at later times (including magnetically driven mass ejection) is not affected by the lack of resolution.
Nontheless, we caution that a future investigation based on a much higher resolution (beyond our present capabilities) will be necessary for an accurate quantitative assessment of the ejecta properties and the associated numerical error. 

The employed numerical codes and setup are those specified in \citet{Ciolfi2020a}. In particular, we recall that the artificial floor density is set to $\rho_\mathrm{atm}\!\simeq\!6.3\times10^{4}$\,g/cm$^3$.
This corresponds to a mass of $\simeq\!2\times10^{-3}\,M_\odot$ within a sphere of radius $2360$\,km, which is the largest distance at which we monitor matter outflows.
The ejecta component of interest carries a mass that is one order of magnitude higher, ensuring that the effects of the floor density could at most represent a small correction. Nontheless, the possibility that this could slightly slow down the ejecta expansion at the largest distances should be taken into account (see discussion in Section~\ref{sec:ejecta}).

%%%%%%%%%%%%%%%%%%%%%%%%%%%%%%%
\section{Analysis of the MNS wind component of the ejecta} 
\label{sec:ejecta}

\noindent The merger product is a magnetized MNS remnant characterized by strong differential rotation \citep{Ciolfi2020a}. The Kelvin-Helmholtz instability in the first few ms after merger (e.g., \citealt{Kiuchi2015,Kiuchi2018}) and the magnetorotational instability later on (e.g., \citealt{Balbus1991,Siegel2013}) amplify the magnetic field by orders of magnitude up to a physical saturation achieved around 50\,ms after merger, corresponding to $E_\mathrm{mag}\!\sim\!10^{51}$\,erg \citep{Ciolfi2020a}.
During this phase, the build-up of magnetic pressure drives a baryon-loaded wind that pollutes the environment around the MNS. Since the overall magnetic field structure is still quite disordered, the mass outflow is nearly isotropic (e.g., \citealt{Siegel2014,Ciolfi2019}). 

As the evolution proceeds, the magnetic field gradually acquires a helical structure along the MNS spin axis that enhances the outwards acceleration in such direction, resulting in a faster component of outflowing material that starts emerging $\approx\!100$\,ms after merger \citep{Ciolfi2020a}.\footnote{
The basic mechanism behind this further acceleration along the spin axis is analogous to the one at play when a strong large-scale dipolar field is superimposed by hand on a nonmagnetized differentially rotating NS \citep{Shibata2011,Kiuchi2012,Siegel2014,Ruiz2018,Moesta2020}. However, while the emergence of a collimated outflow is ubiquitous in the latter setup, the same outcome is not always guaranteed for a more realistic magnetic field evolution through the BNS merger (see discussion in \citealt{Ciolfi2020a,Ciolfi2020b}).
}
As illustrated in Fig.~\ref{fig1}, the faster outflow interacts then with the slower and more isotropic surrounding material, eventually leading to a wide angle ejecta component with a velocity profile that is maximum along the axis and declines at higher polar angles. 
At the end of the simulation, further mass ejection is significantly reduced, while the material that remained bound starts to slowly fall back towards the MNS. 
\begin{figure}[!t]
  \centering
  \includegraphics[width=0.99\linewidth]{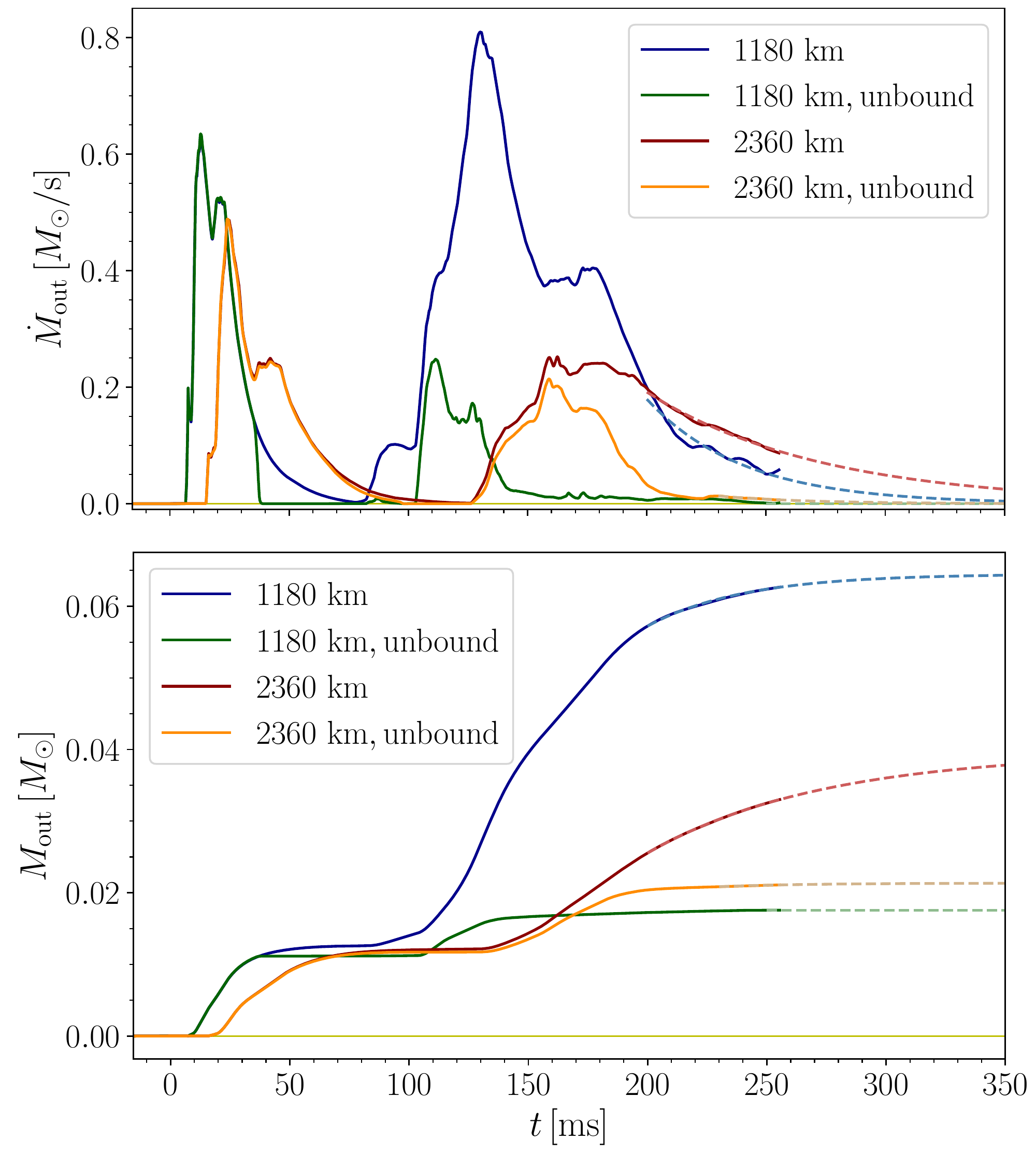}
  \caption{{\it Top:} Time evolution of the mass flow rates across spherical surfaces of radius 1180 and 2360\,km, for both the total mass and the unbound mass. Dashed lines correspond to the exponential profiles used for the extrapolation at later times.
  {\it Bottom:} Analogous plot for the cumulative mass flows.}
  \label{fig2}
\end{figure}
\begin{figure}[!t]
  \centering
  \includegraphics[width=0.983\linewidth]{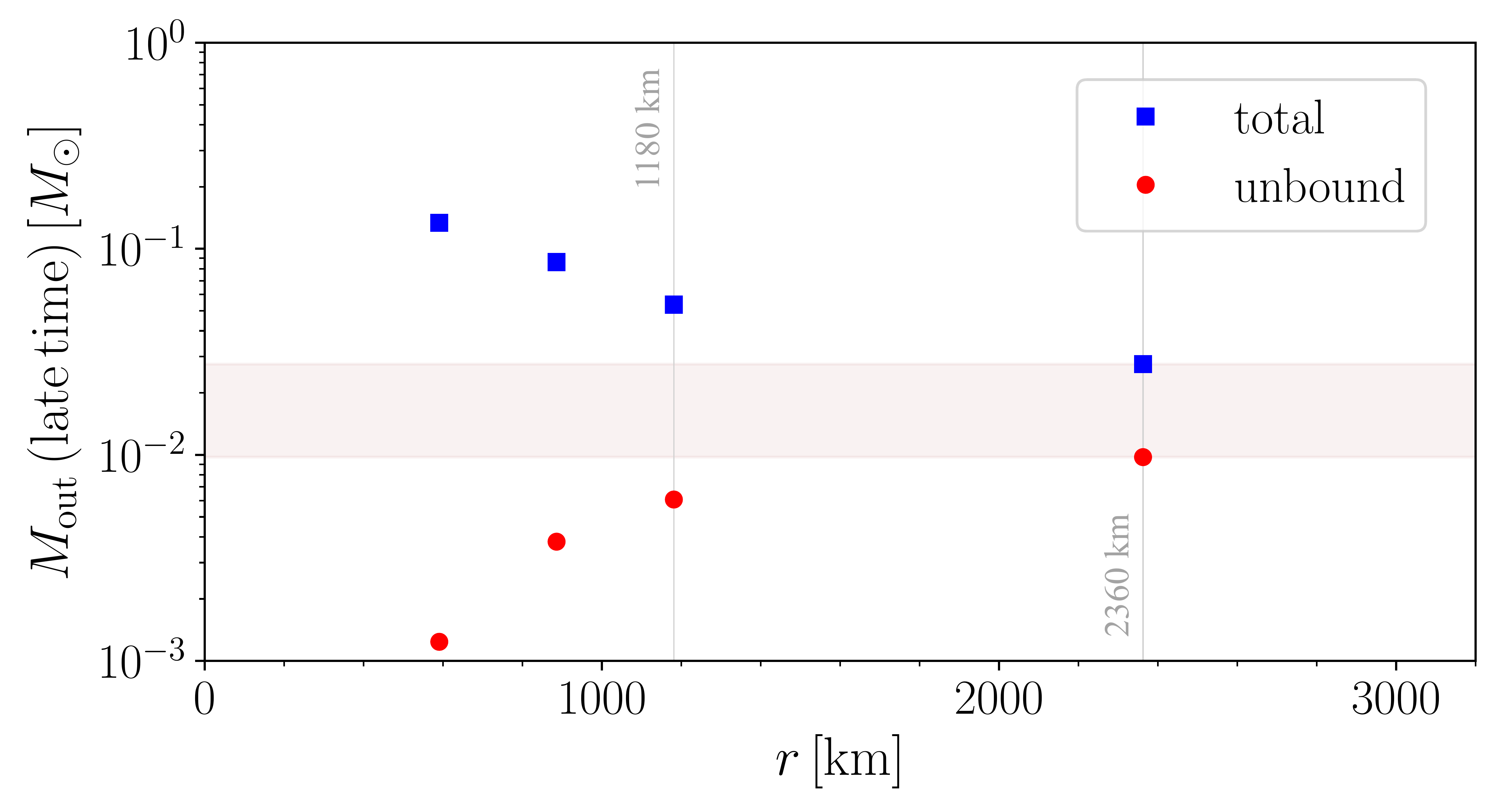}
  \caption{Total and unbound cumulative mass flowing across spherical surfaces of different radii in the late time limit (i.e.~maximum achievable for a sufficiently long-lived MNS). The values refer to the MNS wind only, excluding dynamical ejecta contributions. The ultimate ejecta mass (in the late time and large distance limit) has to be within the range defined by the shaded region.}
  \label{fig3}
\end{figure}

The time evolution of the mass flow rate across spherical surfaces of different radius is shown in Fig.~\ref{fig2}, along with the corresponding cumulative mass. 
In particular, we report the result for 1180\,km and 2360\,km distance and for both the total mass and the unbound mass. To define the unbound mass, we adopt the conservative geodesic criterion $-u_t\!>\!1$ \citep{Hotokezaka2013}.

The early contribution of dynamical ejecta can be clearly distinguished from the MNS wind contribution. We find $\simeq\!0.01\,M_\odot$ of dynamical ejecta expelled within less than 10\,ms after merger (with no distinction of tidal and shock-driven ejecta). Having high enough velocity (ranging up to more than $0.2\,c$), this component could represent a potential candidate to explain the blue kilonova in AT\,2017gfo. However, only a fraction of this material would retain a low enough opacity (e.g., \citealt{Radice2016}) and given that the mass is already smaller than the expected range (i.e. $\simeq\!0.015-0.025\,M_\odot$), the chances of being the dominant source of the blue kilonova appear rather limited. 
While our results do not exclude that dynamical ejecta could contribute significantly to the blue and/or the red kilonova components of AT\,2017gfo, a more precise assessment is beyond the scope of this work.

Turning the attention to the MNS wind, Fig.~\ref{fig2} shows how the (total and unbound) mass flow rate at a given distance grows in time up to a maximum and then starts declining, with a clear suppression towards the end of the simulation ($\simeq\!255$\,ms after merger). 
The final decline is quite regular and can be reproduced very well with an exponential decay. Based on this, we can extrapolate in time to obtain the full evolution. 
At this point, it is possible to compute the total and unbound cumulative masses in the late-time limit (assuming that no collapse to a BH occurs) for a given set of distances. 
The result is given in Fig.~\ref{fig3}. As expected, the total mass decreases with distance, but the unbound fraction increases, converging towards the large distance limit in which the two coincide (i.e.~all the outflowing material far away from the MNS is unbound). 
The ultimate ejecta mass will then be contained within the range of values defined by the total and unbound masses at the largest distance that we probe, 2360\,km, leading to $M_\mathrm{ej,\,wind}\!\simeq\!0.010\!-\!0.028\,M_\odot$. We note that this refers to the MNS wind only, excluding the dynamical ejecta contribution. 

The above mass would be sufficient to explain the blue kilonova in AT\,2017gfo. While limited to a single physical model, this result shows that, in terms of mass, magnetically driven MNS winds represent a valid (and promising) possibility.\footnote{Conversely, explaining the full (blue and red) kilonova mass, which is most likely $\gtrsim\!0.05\,M_\odot$ (e.g., \citealt{Kasen2017,Villar2017}, but see also \citealt{Kawaguchi2018}), seems challenging. This favours the eventual collapse to a BH and the presence of a massive wind from its accretion disk as an additional necessary contribution (in particular for the red kilonova; e.g., \citealt{Siegel2018}).}

The possible collapse to a BH would interrupt the mass ejection process via the MNS wind. When this occurs, only the material within $\approx\!200\!-\!300$\,km from the BH is influenced by the collapse, while the rest is almost unaffected \citep{Ciolfi2019}. Taking our present simulation as a reference, we can then monitor the cumulative mass outflow across the spherical surface of radius $300$\,km and estimate what fraction of mass would actually be expelled within a given collapse time. We find that $90\%$ of the mass outflow (from $5\%$ to $95\%$) occurs between 54 and 189\,ms after merger, while after 285\,ms further outflow accounts for less than 1\% of the total. 
This indicates that a collapse within about $50$\,ms would almost entirely suppress this ejecta component. 
On the other hand, any lifetime beyond $\approx\!200$\,ms would lead to a similar ejecta mass, since after this time additional contributions become negligible.\footnote{At much later time, if no collapse has occurred, the fall-back of material may still lead to additional episodes of mass ejection, but we do not speculate here on whether this is possible and/or quantitatively relevant.
Fall-back material may also leave an imprint in the properties of a later jet (e.g., \citealt{Lamb2020}). }

As shown in Fig.~\ref{fig1}, MNS wind radial velocities reach over $0.2\,c$, in particular along polar directions within $\simeq\!15^\circ$ from the the spin axis. Overall, the velocity of the unbound material spans the range $\simeq\!0.1\!-\!0.2\,c$. These values are consistent with the lower end of the range inferred from observational data (i.e.~$\simeq\!0.2\!-\!0.3\,c$). 
We note, however, that various effects not included in our simulation could further enhance these velocities.
One effect is nuclear recombination of free nucleons into $\alpha$-particles, in which part of the available nuclear binding energy would be ultimately converted into outflowing motion (e.g., \citealt{Siegel2018}). Then, the presence of a relativistic jet launched at later times and drilling through the MNS ejecta (necessary to explain the short gamma-ray burst GRB\,170817A; e.g., \citealt{LVC-GRB,Mooley2018b,Ghirlanda2019}) would also slightly increase the kinetic energy of the latter (to be quantified via actual jet simulations).
Neutrino radiation from the MNS could also lead to a higher radial kinetic energy per baryon within the magnetically driven ejecta (e.g., \citealt{Perego2014}), even though this remains to be demonstrated in a full magnetized BNS merger simulation including neutrino emission and neutrino heating. 
In addition to these effects, we also note that in our simulation an artifical floor density of $\simeq\!6.3\times10^{4}$\,g/cm$^3$ is imposed, which could be slowing down the MNS wind in a non-negligible way (e.g., \citealt{Martin2018}). 

As additional note, the presence of a non-negligible amount of shock-driven dynamical ejecta along the polar direction would imply a reprocessing of the kilonova emission from the MNS wind, which may result in a significant increase of the observed photospheric velocity. This effect, illustrated in \citet{Kawaguchi2018}, represents another viable way to further improve the  accordance between the blue kilonova observations and the somewhat slower MNS wind obtained here.

The angular distribution of the MNS ejecta extends up to a half-opening angle of more than $60^\circ$. The monotonic decrease in velocity from the axis to higher polar angles can be approximately reproduced by $v_r(\theta)=v_r(\theta\!=\!0)/[1+(\theta/
\theta_0)^a]$, where the parameters $(\theta_0$, $a)$ vary significantly in time. For instance, at 120\,ms after merger $\theta_0\!\simeq\!43.5^\circ$, $a\!\simeq\!3.46$ (error $<\!15\%$), while at 180\,ms after merger $\theta_0\!\simeq\!25.8^\circ$, $a\!\simeq\!1.14$ (error $<\!7\%$).

We close the present Section with a note on the opacity.
As confirmed in a number of studies (e.g., \citealt{Perego2014,Fujibayashi2018}), neutrino irradiation (not included here) has the effect of significantly raising the electron fraction of the MNS wind material, leading to ejecta having $Y_\mathrm{e}\!\simeq\!0.25\!-\!0.40$.
The r-process nucleosynthesis is then mostly limited to heavy elements with $A\!\lesssim\!140$ and the resulting opacity remains rather low, i.e.~$\sim\!0.1\!-\!1$\,cm$^2$/g (e.g., \citealt{Tanaka2018}). This is in agreement with the blue kilonova requirement.

%%%%%%%%%%%%%%%%%%%%%%%%%%%%%%%
\section{Conclusions} 
\label{sec:concl}

\noindent We analyzed the outcome of a BNS merger simulation showing the formation of a magnetically driven collimated outflow along the rotation axis of the MNS remnant. 
Such an outflow is associated with the gradual build up of a helical magnetic field structure due to the action of the strong differential rotation within the MNS core.  
We found that the magnetically enhanced mass flow rate and velocity in the polar direction leads to an ejecta component that would represent a viable explanation for the puzzling blue kilonova in AT\,2017gfo.\footnote{ We note that the same ejection mechanism may also offer a viable explanation for other blue kilonova candidates found in association with earlier short GRBs (e.g., \citealt{Troja2018,Lamb2019}). }
In particular, for the case at hand, we obtained a robust estimate for the mass of the unbound material in the range $\simeq\!0.010-0.028\,M_\odot$, with 95\% of the mass ejection occurring within $\simeq\!190$\,ms after merger. This result shows that MNS baryon winds, when a magnetically driven collimated outflow is able to emerge, can lead to an ejecta mass consistent with the blue kilonova observation of August 2017. 
The radial velocities we found are limited to a maximum of $\simeq\!0.2$\,c, consistent with the lower end of the range estimated from the observations. 
We also discuss possible effects that, when taken into account, could further increase the ejecta velocities. 
Finally, we noted that the expected opacities of the MNS wind material are also fully compatible with a blue kilonova. 

Our findings, in combination with those reported in \citet{Ciolfi2020a}, support a scenario for the BNS merger of August 2017 in which (i) the blue kilonova was mainly powered by radioactive decay within the magnetically driven baryon wind from a metastable MNS, possibly with a significant contribution from shock-driven dynamical ejecta, (ii) the eventual collapse of the MNS formed an accreting BH system able to launch the relativistic jet that powered the short gamma-ray burst GRB\,170817A (e.g., \citealt{LVC-GRB,Mooley2018b,Ghirlanda2019}), and (iii) the baryon wind from the accretion disk around the BH provided the dominant contribution to the red kilonova. 

While we refer here to a single merger simulation, a more exhaustive understanding of magnetically driven MNS winds will require the exploration of a large variety of BNS systems and their diverse physical conditions (including different EOS; e.g., \citealt{Ai2020}).
Nonetheless, the present study already provides new important indications on the qualitative properties of this mass ejection channel.
For the first time, we traced the full process, from the initial development of the baryon wind until the suppression of the outgoing mass flow. We found that the wind material is almost entirely expelled within a limited time window, which for the specific case is between about $50$ and $200$\,ms after merger. On the one hand, this implies a lower limit on the MNS lifetime when this ejecta component is present (here $\gtrsim\!50$\,ms). 
On the other hand, it shows that there is a maximum achievable ejecta mass via this channel that remains nearly unchanged for any MNS lifetime above a certain value (here $\approx\!200$\,ms). The time at which mass ejection is mostly over also marks the beginning of the separation between the unbound material moving outwards and the bound material that is starting to fall back towards the central object. 
Furthermore, we found that the ejecta radial velocity is maximum along the MNS spin axis and smoothly decreases for increasing polar angles, up to the full angular extension of more than  $60^\circ$.

Overall, our results confirm the crucial role played by magnetic fields in the post-merger evolution and in particular in the ejection of mass during the MNS phase (see also \citealt{Ciolfi2020b}).
We encourage further investigation on magnetically driven MNS winds via full magnetized BNS merger simulations, possibly including, as a first necessary improvement, neutrino radiation along with finite temperature and composition dependent EOS.

%%%%%%%%%%%%%%%%%%%%%%%%%%%%%%%
\acknowledgments
\noindent We thank Brian D.~Metzger, Wolfgang Kastaun, Albino Perego, and Daniel M.~Siegel for useful comments. 
J.\,V.\,K. kindly acknowledges the CARIPARO Foundation for funding his PhD fellowship within the PhD School in Physics at the University of Padova.
Numerical simulations were performed on the cluster MARCONI at CINECA (Bologna, Italy). We acknowledge a CINECA award under the MoU INAF-CINECA initiative (allocation INA17\_C3A23), for the availability of high performance computing resources and support. In addition, part of the numerical calculations have been made possible through a CINECA-INFN agreement, providing access to further resources (allocation INF19\_teongrav).

%% the bibliography. 
%% To get the citations to show in the compiled file do the following:
%%
%% pdflatex manuscript.tex
%% bibtex manuscript
%% pdflatex manuscript.tex
%% pdflatex manuscript.tex

%\bibliography{refs}{}
%\bibliographystyle{aasjournal}



%%%%%%%%%%%%%%%%%%%%%%%%%%%%%%%
%%%%%%%%%%%%%%%%%%%%%%%%%%%%%%%
\end{document}